\documentclass[a4paper,12pt]{article}

\title{ Space of events and the time observable}
\author{Nikola Buri\' c {\thanks {buric@ipb.ac.rs}} and Slobodan Prvanovi\' c\\
Institute of Physics, University of Belgrade, \\PO Box 68, 11000 Belgrade, Serbia.}

\begin{document}

\maketitle

\begin{abstract}

Quantum mechanical time operator is introduced following  the
parametric formulation of classical mechanics in the extended
phase space. Quantum constraint on the extended quantum system is
defined in analogy to the constraint of the classical extended
system, and is interpreted as the condition defining the  space of
physical events. It is seen that the peculiar properties of the
time observable, otherwise obtained in the models of time
measurement, are of the classical origin, {\it i.e.}, due to the
quantized classical constraint of the parametric Hamiltonian
dynamics.

\end{abstract}

\section{Introduction}

In non-relativistic quantum mechanics, as well as in classical
mechanics,  time is considered as a parameter of the dynamical
orbits, with an arbitrary initial value and an irrelevant global
scale. Consequently, there is no phase space function in the
classical theory and no self-adjoint operator in quantum mechanics
that would correspond to the time as a physical measurable
quantity. Nevertheless, one does measure the duration of various
processes and one does obtain and records information about time
of occurrence of various events. The need to formulate a
consistent theory of time measurement in quantum mechanics on one
hand and the relativistic covariance on the other demand that the
same mathematical objects should be associated with time and space
observables.

The main obstacle  to define an appropriate self-adjoint operator
to represents adequately a measurement of quantum mechanical  time
are known since long ago \cite{Pauli,Holevo}. If the Hamiltonian
of the system, representing its energy, has semi-bounded, bounded
or discrete spectrum, then its canonically conjugated operator can
not
 posses spectral representation in terms of projector measure on
$R$ (PM) like the operators representing other usual observables.
On the other hand, Galileian covariance demands that the operators
representing energy and time must be conjugated. The way out is to
associate the time observable not with PM but with a measure on
$R$ in terms of positive operators that need not be orthogonal on
non-overlapping domains (POVM)\cite{Holevo},\cite{Bush}. This
approach has been used by many to model time observables for
different physical systems \cite{Holevo} or to model particular
measurement schemes corresponding to different notions of time,
such as the time of arrival, the tunneling time and the time of a
quantum clock associated with the phase variable \cite{Muga}.

  Well known
parametric formulation of classical Hamiltonian systems, based on
an extended phase space, is often used to transform an explicitly
time-dependent system into an equivalent autonomous one
\cite{Lanczos,Arnold,Stuckmeier}. In this formalism the time
parameter is transformed into a coordinate of the extended phase
space, and treated (almost) analogously as other canonical
coordinates. The time-coordinate of the extended phase space is
sometimes called the ideal time, or external time,  because it is
meant to be universal and not related to particular type of time
measurement. However, the extended system has an additional
constraint, which introduces  ambiguity into its quantization.
This theme is usually treated under the name of quantization of
parameterized non-relativistic or relativistic particle {a
comprehensive review is presented in \cite{Ruffini}), which are
understood as simplified models with the typical problems that
appear in the quantization of general relativity (a recent review
of "the problem of time" in general relativity can be found in
\cite{Macias},\cite{Kucar}).

On the other hand, and independently of the quantization of the
parametric time, different types of quantum time observables have
been introduced as mathematical models of measurements of
occurrence times of particular events. As we have already stated,
within this research theme an important common property of
different time observables has been established: the time
observables are mathematically represented by POVM's on time
values domains, and not by PM's like most dynamical observables.

The goal of our paper is to show that the main mathematical
property of different time observables, i.e. the POVM
representation, can be seen as a property of the parameter time
variable and its quantized version.
 Essential property, common to the different quantum time observables
  is thus seen as implied already by the classical constraint on the extended
  Hamiltonian system. This indicates deep relation
and unity of these different concepts of the time observables.

In the next section we shall review the parametric formulation of
a Hamiltonian dynamical system. This includes the definition of
the classical time variable in the extended phase space, and the
constrained classical evolution equation. In Section 3 we
introduce the extended quantum system. It should be stressed right
at the beginning that we shall not quantize the constrained
classical extended system. Instead the analog of the classical
constraint is introduced as a condition on physically admissible
states of the extended Hilbert space. We shall see that this
constraint implies properties of the quantum time observable.
Summary of the main argument is presented in Section IV.

\section{Extended phase space and representation of time}

  Well known parametric formulation of Hamiltonian systems via
 the extended symplectic phase space \cite{Lanczos,Arnold,Stuckmeier}
  suggests a way to associate time with an operator defined on an extended Hilbert space.
  The parametric formulation is usually used to transform a
  non-autonomous system into an autonomous one. However, here
  it is used in order to introduce the time variable for an
  autonomous system.
 Consider a Hamiltonian system $(R^{2n},\Omega,H)$, where $\Omega$ is the standard symplectic structure
\begin{equation}
\pmatrix{0&1\cr -1&0},
\end{equation}
  where  $0$ and $1$ are the  $n$ dimensional zero and unit
  matrices,
  and $H$ is the Hamilton's function. Associated with this system
  is an extended system on
 $(R^{2n+2},\Omega_{ex},H_{ex})$ defined as follows. The canonical coordinates of the extended phase space $R^{2n+2}$
 are $q_1,q_2\dots q_{n+1},p_1,p_2\dots p_n,p_{n+1}$ where $q_1,q_2\dots q_n,p_1,p_2\dots p_n$ are the canonical coordinates of the
 original system and $(T\equiv q_{n+1},S\equiv p_{n+1})$ are canonical coordinates in the added symplectic plane. The extended symplectic form
  $\Omega_{ex}$ is of the
 form (1)
  with  $0$ and $1$ being the $n+1$ dimensional zero and unit matrices. The Poisson brackets between the canonical coordinates
 are determined by the extended $\Omega_{ex}$. In particular:
\begin{equation}
\{T,S\}=1,\>\{T,q\}=\{T,p\}=\{S,q\}=\{S,p\}=0.
\end{equation}
Let us mention that we do not  gain anything by treating more
general symplectic manifold instead of $R^{2n}$ since in order to
quantize the classical system, which is to be done in the next
section, canonical coordinates have to be specified in advance.

In order that the two systems $(R^{2n},\Omega,H)$ and
$(R^{2n+2},\Omega_{ex},H_{ex})$ describe the same dynamics the
Hamiltonians $H$ and $H_{ex}$ are related as follows:  Hamilton
variational principle for the original and extended systems are
equivalent if
\begin{equation}
H_{ex}d\theta=(H+S)dt,
\end{equation} where $t$ and $\theta$ are evolution
parameters in the original and in the extended systems. The
original Hamiltonian defines invariant hypersurfaces
 in $R^{2n}$ by $H(q,p)=const$. The equivalent requirement on the extended $H_{ex}$, which need to be determined only up to an additive constant,
suggests to define $H_{ex}$ as an implicit function $H_{ex}=0$.
This now defines a $2n+1$ dimensional hypersurface in $R^{2n+2}$.
In view of the equation
 (3) this choice relates the value of the canonical coordinate $S$ to the value of the original Hamiltonian
\begin{equation}
S(\theta)=-H(q(\theta),p(\theta))\equiv -h(\theta).
\end{equation}
This choice uniquely determines the new Hamiltonian $H_{ex}$ as
\begin{equation}
H_{ex}d\theta=(H-h)dt
\end{equation} so that the Hamilton's variational
principle with the original Hamiltonian $H$ is equivalent to the
extended formulation with the extended
 Hamiltonian and the constraint
\begin{eqnarray}
H_{ex}&=&k[H-h],\> k=dt/d\theta ,\nonumber\\
H_{ex}&=&0.
\end{eqnarray}
The appearance of the scaling factor $k=dt/d\theta$ ensures that
the extended system is covariant with respect to the canonical
transformations which might
 act nontrivially on all extended canonical coordinates including $T$ \cite{Stuckmeier}.
The most common and traditional choice for the value of the
scaling factor $k$ is $k=1$ \cite{Lanczos, Arnold}. In order to
simplify the presentation of the main arguments, in what follows
we shall also always take $k=1$.

 The dynamical equations of the extended system in terms of
the new parameter $\theta$ are of the usual canonical form:
\begin{equation}
{dg_i\over d\theta}={\partial H_{ex}\over \partial p_i}, \quad
{dp_i\over d\theta}=-{\partial H_{ex}\over \partial q_i},\>
i=1,2,\dots n+1,
\end{equation}
and are equivalent to the original equations in terms of the
parameter $t$:
\begin{eqnarray}
{dg_i\over dt}&=&{\partial H\over \partial p_i}, \quad {dp_i\over dt}=-{\partial H\over \partial q_i},\> i=1,2,\dots n,\nonumber\\
{dt\over d\theta} &=&1,\quad {dh\over d\theta}={\partial H\over
\partial t}=0,
\end{eqnarray}
with the assumed relation between the parameters.

The advantage of the extended formulation is that the parameter
time of the original formulation is treated as a canonical
coordinate $T$
 on an equal footing as the other canonical variables that correspond to the spatial degrees of
 freedom. The extended phase space and its submanifold given by the constraint (6) could
 perhaps be called the space of events and the manifold of
 physical events respectively.

 {\it Remarks}

 The following three remarks are not crucial for our main
 argument.

 $1^o$ It is obviously equivalent to the presented formulas to define the extended Hamilton's
 function with $-p_{n+1}$ in the equation (3) instead of
 $+p_{n+1}$, and to accompany this by the same change in
 the corresponding canonical Poisson bracket.

 $2^o$ Going in the direction opposite to the construction of the
extended system one can attempt to introduce an intrinsic time
variable as a suitable function of the $2n$ canonical variables
(see for example \cite{Peres}). Such intrinsic time behaves as the
phase variable of the dynamical system. However, the possibility
to define  a valuable intrinsic time is tightly related to a very
difficult problem of integrability of the dynamical system.

 $3^0$ It is well known that a quantum systems with the Hilbert
 space ${\cal H}$ can be considered as a Hamiltonian dynamical
 system with the projective Hilbert space $P{\cal H}$ as the
 symplectic phase space \cite{Kible,Ashtekar, Hughston}. It might be tempting to try to introduce
 a  quantum mechanical time observable using the parametric formulation of the
  Hamiltonian representation of
  quantum mechanics. In the Hamiltonian formulation
  the real and the imaginary parts of the
 Hermitian scalar product, reduced on $P{\cal H}$,  generate
  the Riemannian and the simplectic structure on the phase space respectively. The
  quantum states are represented by points of the phase space and
  the observables $\hat A$ by functions $<\hat A>$ whose Hamiltonian vector fields
  generate  isometries. The Schroedinger evolution equation is
  reproduced by the Hamilton dynamical equations with the
  Hamilton's function $H=<\hat H>$ where $\hat H$ is the
  Hamiltonian. The function representing the commutator between
  two observables is given by the Poisson bracket of the
  corresponding functions. The geometric formulation of quantum mechanics has been used to study
also the constrained quantum dynamics
\cite{const1,JaAnnPhys,Brody}.
  One could now introduce the parametric
 formulation on the extended phase space of the Hamiltonian system
 corresponding to a quantum system. However, thus introduced
 canonical time coordinate does not define an observable.  Only a small subset of functions on the phase
space $P{\cal H}_{ex}$ correspond to quantum mechanical
observables. In particular the canonical coordinates $(q,p,T,S)$
of a point do not represent observables.  The canonical
coordinates represent components
 of the quantum state in some basis. Thus, the coordinate corresponding to the time in the
extended parameteric formulation does not correspond to an
observable. Several extensions of quantum mechanics
\cite{Ashtekar} that generalize the geometric formulation to
  include the coordinates $(q,p)$ as legitimate observables
 are seen as theories with hidden variables enabling one to uniquely measure and determine the quantum
 state.

\section{Quantum mechanical time observable}

{\it Extended quantum system}

The extended phase space of a classical system is formed by
treating time as an additional independent degree of freedom, and
the parametric dynamics represents a constrained system on the
extended phase space. Quantization of the system (6) is often
discussed as a simple example of constrained parametric system
quantization \cite{Dirak}, \cite{HT}. We shall follow one of the
possible quantization procedures in which the extended phase space
is canonically quantized as if there are no constraints. The
constraints are then included in the form of conditions imposed on
the space of physical states.

 The classical approach suggests an analogous treatment of a quantum system
  in which the system's  Hilbert space ${\cal H}$ is considered as
 the tensor product of the Hilbert spaces ${\cal H}_s$ corresponding to the spatial degrees of freedom and the Hilbert space ${\cal H}_T$
 that corresponds to the time treated as an additional degree of
 freedom. The extended system will be defined by: a) the commutation relations between the time and its
 conjugate, b)
  the extended  Hamiltonian and c) the constraint.
There is an ambiguity in the
 choices of the sign in the commutation relations between the time
 and its conjugate.  However the alternatives are equivalent in the sense that the properties of the physical time observable
 constructed in the two alternative ways are the same.  We shall
 present both alternatives in parallel.

   The extended Hilbert space is defined as
    the direct product ${\cal H}_{ex}={\cal H}_s\otimes {\cal H}_T$.
     The structure of ${\cal H}_T$ is dictated by the desired algebraic properties
 of the time variable represented by an operator $I\otimes \hat T$ acting on ${\cal H}_T$. Pursuing the analogy with the classical extended
 phase space with the Poisson-Lie bracket $\{T,S\}=1$ the Hilbert space ${\cal H}_T$ should carry an irreducible representation of
the same Lie algebra
\begin{equation}
 [\hat T,\pm\hat S]= i.
\end{equation}
 Thus,
 %${\cal H}_T={\cal L}_2(R)$, and
 $\hat T$ and $\hat S$ are
 represented by  multiplication and differentiation operators acting  on functions from the corresponding domains in ${\cal
 H}_T$.
 %={\cal L}_2(R).
Of course, $\hat T$ and its conjugate $\hat S$ commute with
operators acting in ${\cal H}_s$. In particular, $\hat T$ commutes
with
 Hamiltonian operators $\hat H_s$. The $\pm$ sign   will be fixed to correspond to the sign of the operator $\hat S$
in the extend Hamiltonian.

  Operators $\hat T$ and $\hat S$
 have continuous spectra on ${\cal H}_{ex}$. Nevertheless, we shall
 often use the terminology "eigenvalue" and "eigenvectors" for
 those operators indicating by the quotation marks that these
 should be understood in the generalized sense.
 For, example if the common "eigenvectors" of the coordinate
 variables are denoted by $|q>$ and if the "eigenvector" of the
 time operator $\hat T$ is denoted by $|T>$ then
the operator $\int_V|q><q|dq\otimes\int_{\Delta T}|T><T|dT$ is
interpreted as the property that the system is in the volume $V$
during the time interval $\Delta T$.
   Unhated letters like $H_s, T, S,H_{ex}$ denote the
  eigenvalues or "eigenvalues" of the corresponding operators, which are, of
  course,
  to be distinguished from the corresponding classical functions
  denoted by the same symbols in the previous section.

At this stage we can propose that general mixed state of the
extended system are represented by statistical operators
$\rho_{ex}$ in ${\cal H}_{ex}$, and pure states are linear
 combinations of separable states:
 \begin{equation}
  \sum_{i,j}c_{ij}|\psi_i>_s\otimes |\phi_j>_T.
  \end{equation}
 Of course, vectors
 corresponding to pure states that have different norm or
 phase are all supposed to represent the same  state. In what follows we
 shall not explicitly take care about this gauge invariance of the
 states.

 Consider a system with the Hamiltonian $\hat H_s=
\hat H_s(\hat q,\hat p)$.
 In order that the original system on ${\cal H}_s$ and the extended one on ${\cal H}_{ex}$ describe the same quantum evolution the Hamiltonian of the
 extended system is defined as
\begin{equation}
\hat H_{ex}=(\hat H \pm\hat S).
\end{equation}
Notice that $\pm$ signs in the Hamiltonian (11) correspond to
$\pm$ sign in the commutation relation (9), analogously to the
situation in the classical case, eq. (2) and (3).

 Indeed, consider the Schroedinger evolution of an extended separable pure state
 $|\psi >=|\psi>_s\otimes |\psi>_T$
\begin{equation}
i{\partial |\psi >\over \partial\theta}=\hat H_{ex}|\psi >=(\hat
H_s|\psi>_s)\otimes|\psi>_T\pm | \psi>_s\hat S|\psi>_T.
\end{equation}
On the other hand
\begin{equation}
i{\partial |\psi >\over \partial\theta}=i{\partial |\psi>_s\over
\partial t}\otimes |\psi>_T+i|\psi>_s\otimes {\partial |\psi>_T\over
\partial t}.
\end{equation}
Thus
\begin{eqnarray}
i{\partial |\psi>_s\over \partial t}&=&\hat H_s|\psi>_s\\
i{\partial |\psi>_T\over \partial t}&=\pm&\hat S|\psi>_T
\end{eqnarray}
The Schroedinger equation for ${\cal H}_s$ (14) is reproduced, and
the equation (15) determines the operator $\hat S$. We see that
the choice of $\pm$ in the commutation relations (9) has to be
performed together with the corresponding choice in the extended
Hamiltonian (11).

In either case (10) the Hamiltonian $\hat H_{ex}$ and the ideal
time operator $\hat T$ satisfy
\begin{equation}
\Delta  H_{ex}\Delta T\geq<i[\hat H_{ex},\hat T]/2>=1/2.
\end{equation}
Of course, the physical interpretation of  (15) is not that of the
time-energy uncertainty relation, since the system's energy is
represented by the Hamiltonian $\hat H_s$ and commutes with $\hat
T$.

 {\it Constraint}

 The classical constrain $H_{ex}=0$ is introduced into quantum mechanics by a
  constraint which must be satisfied by the states of the extended
 system as follows: It is declared that not all vectors from ${\cal H}_{ex}$
 should be considered as representing states of the physical
 system but only those that satisfy the following condition, analogous to the
 classical equation of the constraint
 \begin{equation}
 \hat H_{ex}|\psi>=(\hat H_s\pm\hat S)|\psi>=0,
 \end{equation}
which  is equivalent to
 \begin{equation}
 <\psi|(\hat H_s\pm\hat S)^2|\psi>=0.
 \end{equation}
Since $\hat H_s$ and $\hat S$ are linear, the set of physical
states is a linear subspace of ${\cal H}_{ex}$. We denote the
space of physical states, {\it i.e.}, those that satisfy (17) by
${\cal H}_{phys}\subset {\cal H}_{ex}$. The space ${\cal
H}_{phys}$ will be called  the space of physical events to
emphasize the fact that ${\cal H}_{phys}$ is a subspace of ${\cal
H}_{ex}$ and not of ${\cal H}_{s}$.

In summary, the evolution equation on ${\cal H}_{ex}$ is given by
the Schroedinger equation with the extended
  Hamiltonian $H_{ex}$, but the physical
pure states are
 only those vectors in ${\cal H}_{ex}$ that satisfy the condition of
 the (quantum ) constraint (17).
 The quantum constraint (17) is
 obviously consistent with the extended dynamical equation.

 The constraint (17) implies that the action of $\hat H_s$ and
 that of $\mp\hat S$ on the vectors from the subspace ${\cal H}_{phys}$
coincide. Consider the vector $|\psi>=|E_i>\otimes |S>$ where
$|E_i>$ is an eigenvector of $\hat H_s$  and $|S>$ is a
 "eigenvector" of $\mp\hat S$. Due to the constraint (17)
 the spectra and the eigenstates of $\mp\hat S$ restricted on
the
 space of physical events ${\cal H}_{phys}$ are equal to the spectra and the eigenstates of $\hat
 H_s$ restricted on ${\cal H}_{phys}$.
The general physical state $|\psi>_{phys}\in {\cal H}_{phys}$ is a
linear combination of the states on the diagonal of ${\cal
H}_{ex}$:
\begin{equation}
|\psi>_{phys}=\sum_ic_i|E_i>\otimes|S_i=E_i>,
 \end{equation}
 where $|S_i=E_i>$ denotes the "eigenstate" of $\hat S$ with the
"eigenvalue" numerically equal to an energy "eigenvalue" $E_i$.
The formula (18) for a physical state is to be compared with the
representation (10) of a general vector in ${\cal H}_{ex}$.

  Notice that the restriction of $\hat S$ has
 the spectra of the restriction of $-\hat H_s$ if the
 commutator $[\hat T,\hat S]=i$ and the spectra of $+\hat H_s$ if
$[\hat T,\hat S]=-i$. Thus, the relation between the algebra of
the operators $\hat T$ and $\hat S$ and  the physical consequences
of the quantum constraint (17), is the same as the relation
between the Poisson algebra of the classical quantities $T,S$ and
the physical consequences of the classical constraint given by
equation (4).

  The restriction of $\hat T$ onto
 the ${\cal H}_{phys}$ is denoted by $\hat T_{phys}$, and could be called the physical
 time. Due to the constraint (17) and the consequent properties of
 $\hat S_{phys}$, we can conclude that $\hat T_{phys}$ does not
 generate an orthogonal resolution of unity. To this end we
 recall the operational treatment of the time
observable\cite{Holevo},\cite{Bush}. In this approach the
canonical commutation relation between an energy operator and a
formal operator representing time is interpreted in mathematically
rigorous way as existence  of a covariant POVM associated to the
time observable.
 The restrictions of $\pm\hat S$ and
 $\hat T$ on the subspace of physical events are in the same
 relation as the energy operator and the formal time operator in the operational approach.
 Thus, we conclude that $\hat T_{phys}$, {\it i.e.}, the restriction of $\hat T$ onto the
 space of physical events ${\cal H}_{phys}$,  generates the corresponding POVM from $R$ onto the subspaces of
 ${\cal H}_{phys}$. Let us stress that the properties of the
 physical subspace ${\cal H}_{phys}$ are determined by the
 Hamiltonian of the system $\hat H_s$, but the fact that $\hat
 T_{\ phys}$ generates a POVM and not a PM is universal for all
 physically plausible Hamiltonian operators.

In conclusion, we see that, due to the quantized classical
constraint (17), there is no physical state where the formal time
observable $\hat T$ has definite sharp value. "Eigenstates" of the
restriction of $\hat T$ on ${\cal H}_{phys}$ are not only
non-normalizable, but are more importantly non-orthogonal.
 Thus, despite the fact that the
operators
 representing the system's energy $\hat H_s$ and the formal time
 $\hat T$ do commute, there is no physical event where the formal time $\hat
 T$, or its restriction $\hat T_{phys}$,
 has definite value and could be measured with an infinite
 precision. Our construction indicates that the main reason for
 this fact is of the classical origin, namely the classical
 constraint (6), with the quantized form (17) that determines the
 space of physical events.

\section{Summary and discussion}

We have explored the consequences of the  possibility to introduce
a time observable into quantum mechanics by following the
procedure of parametric Hamiltonian mechanics on the extended
phase space. The time appears in the extended classical system as
an additional coordinate, but this sets a constraint that must be
satisfied by the extended system. Quantization of the extended
system results with a pair of canonically conjugate self-adjoint
operators on the extended Hilbert space, which can be considered
as an ideal formal time $\hat T$ and its conjugate.
 In order that the time
operator corresponds to the classical external time the analogy of
the classical constraint had to be introduced. The constraint in
quantum mechanics assumes the form of a condition on the states
that are considered as physical, and this constraint is trivially
compatible with the Schroedinger evolution. Thus, the time
observable is represented by the restriction $\hat T_{phys}$ of
the formal time operator $\hat T$ from the extended Hilbert space
${\cal H}_{ex}$ onto the space of physical events ${\cal
H}_{phys}\subset {\cal H}_{ex}$. It is then argued that the
observable time $\hat T_{phys}$ must have non-orthogonal
generalized eigenstates, and must be represented by a POVM. It is
thus seen that the need to represent the time observable by a POVM
can be traced to the constrained on the extended state space, and
is thus of an essentially classical origin. This is the main claim
of our paper.

It would be interesting to use the time observable $\hat
T_{phys}$, as introduced here, to study the process of time
measurement. Such an analyzes should establish relations between
$\hat T_{phys}$ and the POVM's related to particular measurements
of different time observables.

 {\bf Acknowledgments} This work was supported in part by the Ministry
  of Science and Technological Development of the Republic of Serbia, under project No.
  171017 and 171028.

\end{document}